\documentclass[amsmath,prb,showpacs,twocolumn]{revtex4}
\usepackage{graphicx}
\usepackage{amssymb}

\DeclareMathOperator{\sgn}{sgn}
\newcommand{\trace}{\text{Tr}}
\newcommand{\ham}{\mathcal{H}}
\newcommand{\enshift}{\mathcal{E}}
\newcommand{\myvec}[1]{#1}
\newcommand{\selfen}{\Sigma}
\newcommand{\hamsup}[1]{\ham^{#1}}
\newcommand{\ket}[1]{\left \lvert {#1} \right \rangle}
\newcommand{\bra}[1]{\left \langle {#1} \right \rvert}
\newcommand{\bracket}[2]{\left \langle {#1}|{#2} \right \rangle}
\newcommand{\melement}[3]{\bigl \langle {#1} \bigl |{#2} \bigr |{#3} \bigr \rangle}

\begin{document}
\title{Resonance approximation and charge loading/unloading in adiabatic quantum pumping}
\author{Vyacheslavs Kashcheyevs}
\author{Amnon Aharony}
\author{Ora Entin-Wohlman}
\affiliation{School of Physics and Astronomy,
             Raymond and Beverly Sackler
             faculty of Exact Sciences, \\
             Tel Aviv University, Tel Aviv 69978, Israel}
\begin{abstract}
Quantum pumping through mesoscopic quantum dots is known to be
enhanced by resonant transmission. The pumped charge is close to
an integer number of electrons when the pumping contour surrounds
a resonance, but the transmission remains small on the contour.
For non-interacting electrons, we give a quantitative account of
the detailed exchange of electrons between the dot and the leads
(to the electron reservoirs) during a pumping cycle. Near isolated
distinct resonances, we use approximate Breit-Wigner expressions
for the dot's Green function to discuss the loading/unloading
picture of the pumping: the fractional charge exchanged between
the dot and each lead through a single resonance point is related
to the relative couplings of the dot and the leads at this
resonance. If each resonance point along the pumping contour is
dominated by the coupling to a single lead (which also implies a
very small transmission), then the crossing of each such resonance
results in a single electron exchange between the dot and that
lead, ending up with a net quantized charge. When the resonance
approximation is valid, the fractional charges can also be
extracted from the peaks of the transmissions between the various
leads.
\end{abstract}

\maketitle

\section{Introduction}

There has been much recent
experimental\cite{Switkes01,Kouwenhoven91,Kautz99} and theoretical
\cite{Aleiner98,Brouwer98,Zhou99,
Altshuler99,Shutenko00,WeiWangGuo00,MoskaletsButtikerINELAST01,Levinson02PhA,
Avron01,Makhlin01,WangWangGuo02,Entin02form,Entin02res,Sharma03}
interest in adiabatic quantum pumping through mesoscopic
electronic devices, such as quantum channels or quantum dots
(QDs). Typically, the QD is connected via leads to several
electron reservoirs, and is subject to a slowly varying
oscillating potential, with period $\textsf{T}=2\pi/\omega$. Under
appropriate conditions, the device yields a non-zero dc
time-averaged current between pairs of terminals, even when the
terminals have the same chemical potential.  Under ideal
conditions, the charge $Q$ transferred between the terminals
during a period $\textsf{T}$ may be `quantized', i.e. very close
to an integer times the electron charge $e$. Several recent
theoretical studies have considered enhancement of the adiabatic
pumping current due to resonant transmission through the QD, both
for
non-interacting\cite{WeiWangGuo00,Levinson02PhA,Entin02res,WangWangSC02,BlaauboerSC02}
and interacting electrons\cite{BlaauboerCB01}. Connections between
pumped charge quantization and resonant transmission have been
reported in different contexts\cite{Levinson02PhA,Entin02res,
WangWangSC02,BlaauboerSC02,Aono03,Banerjee03}.

Usually, the oscillating potential is characterized by several
time dependent parameters, $\{X_i(t)\}$. As time evolves during
one period $\textsf{T}$, these parameters follow a closed contour
in the parameter space.
\begin{figure}[tb]
  \includegraphics[clip=true,width=6cm]{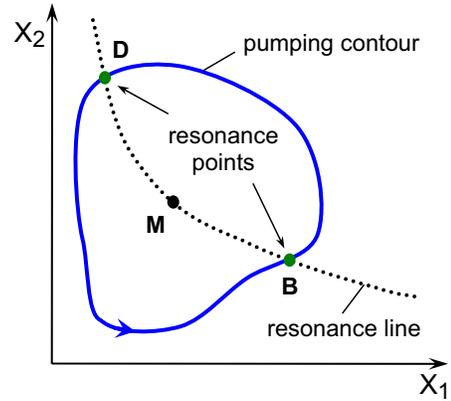}
  \caption{(color online) Schematic picture of
   a two-dimensional pumping contour, crossing the resonance line at two resonance points ($B$ and $D$).
   The transmission is maximal at the point $M$. \label{fig1}}
\end{figure}
A schematic example is shown in Fig.~\ref{fig1}, for two such
parameters.  In parallel to discussing pumping, one can also
consider the conductance between pairs of terminals, generated by
an appropriate bias. This conductance, which depends on the
parameters $\{X_i\}$, may have resonance peaks in the same
parameter space. In this context, one freezes the time dependence,
and considers the conductance at some instantaneous values of the
$\{X_i\}$'s. It has been argued\cite{Levinson02PhA} that the
pumped charge $Q$ will be close to being quantized if the pumping
contour surrounds such a peak (e.g. at the point $M$ in
Fig.~\ref{fig1}), while staying at points with a low conductance.

In the present paper we present an approximate theory for
adiabatic pumping of coherent non-interacting spinless electrons,
which is valid for discrete and distinct resonances, and use this
approximation to obtain physical insight into the reasons for this
quantization. Given a conductance peak (e.g. at the point $M$ in
Fig.~\ref{fig1}), one can usually also identify a `resonance
line', along which the conductance decreases from its peak more
slowly than along other directions\cite{Levinson02PhA,Entin02res}.
Such a line is illustrated by the dashed line in Fig.~\ref{fig1}.
In the example shown in this figure, the resonance line is crossed
by the pumping contour twice, at points $B$ and $D$. Measuring the
instantaneous biased conductance between the two relevant
terminals for each time $t$ during the oscillation period, one
expects two local peaks at these two `resonance points'. Under
appropriate conditions, which include the limit of weak
QD-terminal coupling,  most of the pumped current arises when the
parameters are close to these resonance points: for example, one
can identify a `loading' of the QD by some charge $\Delta
Q_\alpha^{\rm res}$, coming from terminal $\alpha$, at the point
$B$, and an `unloading' of the QD, by $\Delta Q_{\alpha'}^{\rm
res}$, into terminal $\alpha'$, at the point $D$. The resulting
total pumped charge per period approaches a robust,
detail-independent value $Q^R$, which is determined {\emph {only}}
by the ratios of the coupling strengths between the QD and the
different reservoirs at the resonance points. We also show that
$Q^R$ can be related quantitatively to the measured values of the
peak conductances.  $Q^R$ is (almost) quantized (in units of $e$)
when there is one dominant coupling for each resonance.

Our results can be summarized in a very simple and physically
transparent way, by considering the occupation numbers of the
quasi-bound state on the QD, corresponding to each transmission
resonance. Each time the energy of such a state crosses the
chemical potential $\mu$ (which is the same in all reservoirs),
the QD gains or loses one electron, so that the total pumped
charge flowing into it (per period) is quantized. However, the
distribution of the pumped charge between different reservoirs is
proportional to the corresponding coupling strengths (tunnelling
rates). Therefore the pumping current between any two leads can be
obtained by summing up individual resonance contributions, with
appropriate signs.

A similar `shuttling mechanism' for pumping has been used widely
to interpret experiments\cite{Kouwenhoven91,Kautz99} in the
Coulomb blockade regime, when the energetics on the QD is
dominated by the electron-electron
interactions\cite{AverinLihkarev91}. In that approach, electrons
are transferred from a lead to the dot and then from the dot to
another lead, whenever such transfers are favored energetically.
In contrast, Refs.
\onlinecite{Levinson02PhA,Entin02res,Entin02form,Aharony02PRL,Entin03japan}
presented explicit quantum mechanical calculations for pumping of
non-interacting electrons, calculated the total charge pumped
during a full cycle and emphasized the role played by quantum
interference in such processes. In some sense, the present paper
bridges between these points of view: in the limit of weak
coupling between the QD and the leads, we do end up with a
loading/unloading picture, even for non-interacting
electrons\footnote{Distinct charge steps during the pumping cycle
have been presented in Ref.~\onlinecite{Aharony02PRL}, but with no
detailed quantitative discussion of their relation to
resonances.}. However, the details of the charge exchanges during
a pumping cycle are found to be more complicated than in the
`shuttling' picture: at a given resonance point, charge can
usually be shared by several leads. Apart from this, the
conditions for the applicability of our loading/unloading picture
are similar to those of a single electron
transistor\cite{Kouwenhoven91}, in the sense that the role of
quantum interference is restricted to the definition of
independent single-particle resonances. In view of this, there is
room to conjecture that some of our results may also apply in the
presence of electron interactions.

The paper is organized as follows.  In Sec.~\ref{SecGeneral} we
review the physical assumptions of the model and  the formulae
used for the calculation of the adiabatic current. We then use
these formulae to derive the current for a single resonant state,
by approximating the Green function on the QD by a
Breit-Wigner-type formula.  In Sec.~\ref{SecQuantBig} we obtain
our main result --- the adiabatically pumped charge for a sequence
of well-defined distinct resonances --- and discuss possible
applications and experimental verification. To demonstrate this
general picture, Sec.~\ref{secEx} presents the analysis of the
pumped charge for a simple model \cite{Entin02form} of a
``turnstile'' pumping device. A short summary concludes the paper,
in Sec.~\ref{secConc}.

\section{Adiabatic current}\label{SecGeneral}
We consider a spatially confined nanostructure (the QD) connected
by ideal leads to the electronic reservoirs with a common chemical
potential $\mu$ and temperature $T$. The total Hamiltonian for
non-interacting spinless electrons is
\begin{align}
  \ham & = \hamsup{d} +\sum_\alpha \left ( \hamsup{l}_\alpha +
  L_\alpha + L_\alpha^\dagger \right ) \label{labelHamfull} \, , & \\
  \hamsup{d} & = \sum_{mn} h_{mn}(t) \ d^{\dagger}_m d_n \, , & \text{(dot)}  \label{Hdot} \\
  \hamsup{l}_\alpha & =\sum_{\myvec{k}} E_{\alpha\myvec{k}} \ c^{\dagger}_{\alpha \myvec{k}}
  c_{\alpha \myvec{k}} \, , & \text{(leads)} \\
  L_\alpha  & = \lambda_\alpha(t) \sum_{\myvec{k},n}
  J_{\alpha \myvec{k} n} \, c^{\dagger}_{\alpha \myvec{k}} d_n \, .
  & \text{(hopping)}
\end{align}
Here $\hamsup{d}$ is the Hamiltonian of an $N$-state isolated QD
 ($n,m=1,\ldots,N$), the index $\alpha=1,\ldots,L$ enumerates
the  one-dimensional leads connected to  the QD,
$c^{\dagger}_{\alpha \myvec{k}}$ creates a standing wave
$\ket{w_{\myvec{k}\alpha}}$ with wavenumber $\myvec{k}$ and energy
$E_{\alpha\myvec{k}}$ in the channel $\alpha$,  the operator
$L_\alpha$ describes hopping from  the QD into the channel
$\alpha$, and the $\lambda_\alpha$'s are real dimensionless
coefficients. For pumping we allow variation of $\ham(t)$ via the
time-dependent parameters $h_{mn}$ and $\lambda_\alpha$.

The instantaneous adiabatic current in the channel $\alpha$,
directed from a remote reservoir towards the QD, has been
expressed in Ref.~\onlinecite{Entin02form} as
\begin{gather}
  I_{\alpha}(t)  = \frac{e}{2\pi}\int   d  E\, f'(E) \, \mathcal{I}_\alpha,\
  \ \mathcal{I}_\alpha=\frac{1}{\hbar} \,
  \melement{\chi_{k\alpha}}{\dot{\ham}}{\chi_{\myvec{k}\alpha}}
  \,  , \label{pump_current}
\end{gather}
where $f(E)=1/[1+e^{(E-\mu)/k_B T}]$ is the Fermi-Dirac
distribution and $\ket{\chi_{\myvec{k}\alpha}}$ is the
instantaneous scattering state normalized to a unit flux,
$\bracket{\chi_{\myvec{k}\alpha}}{\chi_{\myvec{k}'\alpha}}=
(2\pi/v_{\myvec{k}\alpha}) \, \delta(\myvec{k}-\myvec{k}')$, with
$v_{\myvec{k}\alpha} = \partial E_{\myvec{k}\alpha}/\partial(\hbar
\myvec{k})$ being the velocity in the channel $\alpha$.

In Appendix A we use standard scattering theory formulae to
rewrite this equation in the form
\begin{multline}\label{currentGeneral}
  \mathcal{I}_\alpha(E,t)=\trace_{d}
  \Bigl [ G_d^\dagger \bigl (\dot{\hamsup{d}}+ \dot{\hat{\enshift}} \bigr )
  G_d \hat{\Gamma}_\alpha
  + \bigl ( G_d + G_d^{\dagger} \bigr  )
  \dot{\hat{\Gamma}}_\alpha /2 \Bigr ] \, .
\end{multline}
Here, the operators
\begin{align}
  G_d & = \bigl( E + i 0 -\hamsup{d} - \hat \selfen \bigr )^{-1} \, , \label{Gdot} \\
  \hat\selfen & =  \hat{\enshift}-i \hat{\Gamma}/2 \, ,
  \label{selfenDef} \\
  \hat{\Gamma} & =\sum_{\alpha} \hat{\Gamma}_{\alpha},\ \ \ \
  \hat{\Gamma}_{\alpha} = i L_{\alpha}^{\dagger} (G^{l}_{\alpha}-G^{l\dagger}_{\alpha})
  L_{\alpha} \, , \\
 \hat{\enshift} & =\sum_{\alpha} \hat{\enshift}_{\alpha},\ \ \ \
 2 \, \hat{\enshift}_{\alpha} = L_{\alpha}^{\dagger} (G^{l}_{\alpha} + G^{l\dagger}_{\alpha}) L_{\alpha}
\end{align}
act only on the subspace of the QD. Also, $G^{l}_\alpha$ denotes
the retarded Green's function of an isolated channel,
$G^{l}_\alpha=(E+i0-\hamsup{l}_\alpha)^{-1}$. We have separated
the self-energy operator $\hat\selfen$ into a sum of resonance
width and shift operators\cite{Datta97}, $\hat{\Gamma}_{\alpha}$
and $\hat{\enshift}_{\alpha}$, which are Hermitian.

Equation \eqref{currentGeneral} is a generalized version of the
pumping current formula derived in Ref.~\onlinecite{Aharony02PRL}
for a particular case of single mode tight-binding (TB) leads and
time-independent couplings.\footnote{ Eq.~\eqref{currentGeneral}
can be used to demonstrate the equivalence between the formalism
used in
Refs.~\onlinecite{Entin02form,Aharony02PRL,Entin02res,Entin03japan}
and the adiabatic limit of the Keldysh nonequilibrium Green's
function techniques\cite{JauhoMeir94,WangWangKO02}. Without loss
of generality one can set $\dot{\lambda}_\alpha=0$, and use the
cyclic property of the trace in Eq.~\eqref{currentGeneral}.
Extending the trace to the whole system  and multiplying the
result by two to restore spin-degeneracy gives the charge injected
into the system during an infinitesimal time $\delta t$ as $\delta
Q_\alpha=-(e/\pi) \int d E f'(E) \text{Tr}[G^{\dagger}
\hat{\Gamma}_\alpha G \dot{\ham}\delta t] $. This is identical to
Eq.~(4) of Ref.~\onlinecite{WangWangKO02}, which in turn was
proved\cite{WangWangKO02,WangWang99} equivalent to the Brouwer
formula\cite{Brouwer98,BPT94}.} We next use the general result
\eqref{currentGeneral} for the special case when the pumping is
dominated by a single resonance on the QD. Specifically, we
discuss localization of the accumulated charge and show that
variations of  the charge density in the leads can be neglected
for a thermally broadened resonance (Sec.~\ref{secMasterEq}).

\subsection{Pumping current due to a single level}\label{secResApprox}

The adiabatic current \eqref{currentGeneral} can be calculated
exactly, provided that one is able to compute the Green's function
\eqref{Gdot} on the QD. We are interested in the regime when the
transport is dominated by a single non-degenerate orbital state,
as realized, for example, in a strongly-pinched
QD\cite{Levinson02PhA,Alhassid00}. In this case the transitions
between different energy levels of $\hamsup{d}$ can be completely
disregarded, and instead of  the general Eq.~\eqref{Hdot} it is
sufficient to consider
\begin{align} \label{HdotRes}
  \hamsup{d}_\text{res} & =\epsilon(t) \ket{\psi} \bra{\psi} \equiv \epsilon(t)d^\dagger d \, .
\end{align}
The energy distance to the next resonant state, $\Delta$, will be
assumed to be much larger than all other energies. $\Delta$ can be
set either by the non-interacting effective level spacing of an
arbitrary $\hamsup{d}$ or by the Coulomb blockade away from the
Kondo regime\cite{WangWangKO02}, as discussed in
Ref.~\onlinecite{Levinson02PhA}. In the former case, for example,
our approximation can be applied to the much studied double-delta
barrier pumping
model\cite{WeiWangGuo00,WangWangGuo02,MoskaletsButtiker02} in the
resonant tunnelling regime\cite{WeiWangGuo00,Levinson02PhA}. The
energy level $\epsilon(t)$ may be different at different stages of
the pumping contour, as discussed later on in
Sec.~\ref{SecQuantBig}.

The  Green's function  corresponding to $\hamsup{d}_\text{res}$
now assumes the Breit-Wigner-like form \cite{Breit36}
\begin{align}\label{BWgreen}
  G_d & = \frac{\ket{\psi} \bra{\psi}}{E-\epsilon - \melement{\psi}{\hat{\selfen}}{\psi}} \, .
\end{align}
Substitution of Eq.~\eqref{BWgreen} into
Eq.~\eqref{currentGeneral} gives
\begin{align} \label{ResonanceCurrent}
    \mathcal{I}_\alpha & =  \frac{
   \Gamma_\alpha \, \dot{E}_0 - \dot{\Gamma}_\alpha \, (E_0-E)}
   {(E-E_0)^2 + (\Gamma/2)^2} \, ,
\end{align}
where $E_0(E,t)=\epsilon + \melement{\psi}{\hat{\enshift}}{\psi}$
and
$\Gamma_\alpha(E,t)=\melement{\psi}{\hat{\Gamma}_\alpha}{\psi}$.
Since the partial `width' $\Gamma_\alpha$ is of order
$\lambda_\alpha^2|J_{\alpha k}|^2$, it represents a measure for
the coupling of the QD with the channel $\alpha$. The exact
adiabatic current for a single level given by
Eq.~\eqref{ResonanceCurrent} will be the starting point for our
analysis of the pumped charge in Sec.~\ref{SecQuantBig}.
Breit-Wigner-type expressions for the  current pumped by a single
orbital level have been derived previously in the weak pumping
limit\cite{BlaauboerCB01}, and in the presence of interactions and
Zeeman splitting \cite{Aono03}. However, they were not used to
discuss the details of the pumped charge quantization.

In the remainder of this section we discuss the physical
interpretation of Eq.~\eqref{ResonanceCurrent}. The total current,
$I=\sum_\alpha I_\alpha$, represents changes in the total charge
accumulated both on the dot and in the leads. For small dot-lead
couplings, one would expect that the charge on the QD itself is a
well-defined quantity and a simple picture of single electrons
tunnelling between the leads and the QD should apply. In order to
clarify the relation between our quantum calculation and this
`classical shuttling picture', we first discuss the localization
of the charge and then (Sec.~\ref{secMasterEq}) use a standard
master equation approach to re-derive the current formula in the
limit of $\Gamma \ll  kT$.

Explicitly, Eq. \eqref{ResonanceCurrent} implies that the total
current in our model is a full time derivative,
$I=dQ^\text{F}(E_0,\Gamma)/dt$ of some time-dependent charge
$Q^\text{F}(t)$, where
\begin{align}\label{Qfull}
  Q^\text{F}(t) &= - e \int dE \, f'(E) \left \{ \frac{1}{2} + \frac{1}{\pi}
  \arctan \frac{2(E-E_0)}{\Gamma} \right \} \, .
\end{align}
(We have  chosen the integration constant such that $Q^\text{F}/e$
is bounded between 0 and 1.) The charge $ Q^\text{F}$ can be
interpreted\cite{Klein66,WangWangKO02} as the additional charge
induced in the system by an extra impurity state $\ket{\psi}$, in
the same way as in the Friedel sum rule for the Anderson impurity
model\cite{Langreth66,Klein66}.

This delocalized charge, $Q^\text{F}$, is to be compared with the
local equilibrium occupation inside the QD, which is given by
$Q^\text{occ}/e= \, \trace[\rho \ket{\psi}\bra{\psi}]$, where
$\rho=h^{-1} \int dE \, f(E) \sum_\alpha
\ket{\chi_{\myvec{k}\alpha}} \bra{\chi_{\myvec{k}\alpha}}$ is the
equilibrium density matrix corresponding to
$\ham(t)$.\footnote{The use of the equilibrium $\rho$ is justified
because adiabatic pumping is invariant to the rescaling of time
and can always be performed quasi-statically.}
Using Eqs.~\eqref{LippmanSchwinger}, \eqref{gammadensity},
and \eqref{BWgreen} one can show that
\begin{align} \label{Qd}
  Q^\text{occ} & = \frac{e}{2 \pi} \int dE \, f(E) \, \frac{\Gamma}{(E_0-E)^2 + (\Gamma/2)^2} \, .
\end{align}
If $E_0$ and $\Gamma$ were independent of $E$, then
integration by parts would yield the equality $Q^\text{occ} =
Q^\text{F}$. In general, $E_0$ and $\Gamma$ do depend on $E$, and
hence $Q^\text{occ} \not= Q^\text{F}$.

\subsection{Alternative derivation in the limit  $\Gamma \ll k T$}\label{secMasterEq}

In the limit $\Gamma \ll k T$,  the interference effects in the
leads are expected to be
irrelevant\cite{AverinLihkarev91,Alhassid00} and the resonant
energy level $E_0(t)$ is characterized by its non-stationary
occupation probability $P(t)$, which changes due to tunnelling.
Following Ref.~\onlinecite{Beenakker91}, we assume energy
conservation in the tunnel process and write down the total
tunnelling current in the direction from the lead $\alpha$ into
the QD as
\begin{align} \label{tunnelCurrent}
  I_\alpha =e (\Gamma_\alpha/\hbar) \left \{ [1-P] f(E_0) - P \left [1-f(E_0) \right ] \right \} \, .
\end{align}
The tunnelling rates $\Gamma_\alpha/\hbar$ should be calculated at
$E=E_0$.

In equilibrium, the occupation probability $P_\text{eq}$ is given
by the Gibbs distribution in the grand canonical ensemble
\cite{Beenakker91}. For a single energy level it is the same as
the Fermi-Dirac distribution, $P_\text{eq}=f(E_0)$. Substitution
of $P=P_\text{eq}$ into Eq.~\eqref{tunnelCurrent} gives  a
vanishing net current, as anticipated from the detailed balance.

The kinetic equation for the non-equilibrium distribution
is\cite{Beenakker91,Alhassid00}
\begin{align}\label{Master}
  \dot{P}  & = \sum_\alpha I_{\alpha} /e = (\Gamma / \hbar) [  f(E_0) - P] \, .
\end{align}
For slowly varying $\Gamma$, $E_0$, the time-dependent probability
$P$ will relax quickly towards $P_\text{eq}$. To get an adiabatic
solution, we substitute $P=P_\text{eq} +\tilde{P}$ into
Eq.~\eqref{Master} and set $d \tilde{P}/d t=0$. The result is
\begin{align}\label{adiabaticProbability}
  \tilde{P}  & =  - \hbar f'(E_0) \dot{E_0}/\Gamma \, .
\end{align}
The current \eqref{tunnelCurrent} corresponding to the adiabatic
solution \eqref{adiabaticProbability} is
\begin{align} \label{currentSmallG}
  I_{\alpha} = e (\Gamma_\alpha/\Gamma) f'(E_0) \dot{E}_0
  = e (\Gamma_\alpha/\Gamma) \dot{P}_\text{eq} \, .
\end{align}
(Non-adiabatic corrections can be also included, using the exact
solution of Eq.~\eqref{Master}\cite{LiuNiu93}).

Now we can compare the current derived for $kT \ll \Gamma$,
Eq.~\eqref{currentSmallG}, to the general expression,
Eq.~\eqref{ResonanceCurrent}, which is valid for arbitrary $kT$
and $\Gamma$. By considering $\Gamma/kT$ as a small parameter, one
can perform the energy integration in Eq.~\eqref{pump_current}
with $\mathcal{I}_\alpha$ given by Eq.~\eqref{ResonanceCurrent} to
obtain
\begin{gather}
   I_{\alpha}  = e (\Gamma_\alpha/\Gamma) f'(E_0) \left [ \dot{E}_0 +
   \dot{E}_0 \, O(\Gamma/kT) + \dot{\Gamma}_{\alpha} \, O(\Gamma/kT) \right ]
   \, ,
\end{gather}
which is consistent with Eq.~\eqref{currentSmallG}. The smallness
of the term with $\dot{\Gamma}_\alpha$ comes from the fact that it
is multiplied in Eq.~\eqref{ResonanceCurrent} by an odd function
of $(E_0-E)$.  Similarly, the integrals in Eqs.~\eqref{Qfull} and
\eqref{Qd} give the leading order equalities $Q^\text{occ}  =
Q^\text{F} = e P_\text{eq}$.

We conclude that in the case of a thermally broadened resonance,
the simple tunnelling picture which considers a well-defined
charge on the QD is consistent with the general scattering
approach leading to Eq.~\eqref{ResonanceCurrent}.

\section{Resonance approximation}\label{SecQuantBig}

The Breit-Wigner form \eqref{ResonanceCurrent} of the pumping
current demonstrates a well-established
fact\cite{WeiWangGuo00,Levinson02PhA,Entin02res}, that pumping is
greatly enhanced near a resonance. The resonance condition is
$|E_0-\mu| \lesssim D$, where $D=\max (\Gamma, kT)$ is the
energetic width of the resonance. One option, considered in
Ref.~\onlinecite{BlaauboerCB01}, is to design the pumping contour
in such a way that the system stays entirely at resonant
transmission. In this case,  the Breit-Wigner approximation  does
not lead to any pumped charge quantization\cite{BlaauboerCB01}.
Here we focus on a more generic case, when the resonance condition
is satisfied only during a small fraction of the pumping cycle, as
the system goes through a resonance point.  As shown in Refs.
\onlinecite{Levinson02PhA,Entin02res,WangWangSC02,BlaauboerSC02},
this situation allows for pumped charge quantization.
Specifically, we assume that the system remains near a resonance
point only during a small fraction of the pumping cycle. This
requires relatively narrow resonances, i. e. small widths $D$ and
therefore also small $\Gamma$.

Consider a  resonance time $t_R$ on the pumping contour,
identified by the resonance condition $E_0(\mu,t_R)=\mu$. This
identifies a `resonance point' on the contour. Assume also that
the system 'crosses' this resonance point completely between
times $t_1$ and $t_2$, such that
\begin{enumerate}
\item $\Gamma_\alpha$, $E_0$ are energy independent around the
Fermi surface (for $|E-\mu| \lesssim kT$);

\item  at the `boundary' times, the system is far from the
resonance, $D \ll | E_0(\mu, t_{1,2})- \mu | \ll \Delta$;

\item while at resonance, the couplings change negligibly,
$|\dot{\Gamma}_\alpha| \ll |\dot{E_0}|$.
\end{enumerate}
Under these conditions, we can integrate
Eq.~\eqref{ResonanceCurrent} and get the charge transferred from
the reservoir $\alpha$ in a simple form:
\begin{align} \label{partialQres}
  \Delta Q^\text{res}_\alpha = \int_{t_1}^{t_2} dt \, I_\alpha= -
  e \frac{\Gamma_\alpha}{\Gamma} \sgn \dot{E}_0 \, , \quad
  (\text{at } E_0=\mu) .
\end{align}
For this particular resonance point, other parts of the pumping
contour contribute negligibly to this charge. Equation
\eqref{partialQres} is our main result for the pumped charge due
to a well-defined resonance point.  We will refer to this result
as `the resonance approximation'.  In this approximation, each
reservoir contributes on average a fraction of the electronic
charge, which is proportional to the corresponding  fractional
decay width  or coupling $\Gamma_\alpha/\Gamma$. The total change
in the charge accumulated in the system due to this particular
resonance is thus
\begin{align}\label{totalQres}
  \Delta Q^\text{res} \equiv \sum_\alpha \Delta Q^\text{res}_\alpha=\pm e \, .
\end{align}

This result can be easily generalized for several independent
resonance  points. If the pumping contour can be separated
into  several parts, each containing a single
well-defined resonance point, and  if the
pumping currents on the rest of the contour remain
negligible, then the total charge $Q_\alpha^\text{R}$, pumped
through the channel $\alpha$, is given by a sum over the
resonances: $Q_\alpha^\text{R}=\sum_\text{res} \Delta
Q^\text{res}_\alpha$. For a periodic $\ham(t)$, the pumping
contour is closed, and charge conservation $\sum_\alpha
Q_\alpha^\text{R}=0$ is ensured by Eq.~\eqref{totalQres} and the
fact that the number of loading ($\dot{E}_0 <0$) and unloading
($\dot{E}_0 >0$) resonance  points is the same.

\subsection{Pumped charge quantization} \label{SecQuant}

Equation \eqref{totalQres} can be interpreted as the
loading/unloading of exactly one electron into/out of the QD,
depending on the sign of $\dot{E}_0$ at the Fermi level.
Furthermore, Eq.~\eqref{partialQres} implies that $\Delta
Q_\text{res}$ is dominated by the current from a single channel
$\alpha$, provided that $\Gamma_\alpha \gg \Gamma_{\alpha'}$ for
$\alpha \not=\alpha'$. If the same applies to all the resonances,
then we end up with a ``classical'' picture, in which the pumping
cycle contains a sequence of individual discrete events,  of
exchanging electrons one by one between a reservoir and the QD.
After a full cycle, the charge on the QD will remain unchanged,
and an integer number of electrons will have crossed the QD
between any pair of reservoirs. This gives a detailed explanation
of the pumped charge quantization within this approximation.

Using the same conditions as used to derive
Eq.~\eqref{partialQres}, one can show that both $\Delta Q^F \equiv
Q^F(t_2)-Q^F(t_1)$ and $\Delta Q^{\text{occ}}$ are equal to
$\Delta Q^{\text{res}}$. This means that every time the system
crosses a resonance point, the charge associated with the resonant
state changes by $\sim \pm e$. Therefore we stress that if one  is
interested in the total charge pumped by a single resonance (and
not, for example, in the lineshape of the current,
Eq.~\eqref{ResonanceCurrent}), then the simple picture of
loading/unloading of a single electron, as reflected in
Eq.~\eqref{partialQres}, is applicable  --- regardless of the
ratio $\Gamma/kT$.

We also note that for such an ideal quantization
($Q_\alpha^\text{R} \to e \times \text{integer}$), that is
independent of the contour details, one would need to consider the
limit $\Gamma_\alpha \approx \Gamma \to 0$ for each resonance; the
resonance approximation becomes exact, with results which are
independent of the details of the contour, when $\Gamma
\rightarrow 0$, and the charge goes only via channel $\alpha$ when
$\Gamma_\alpha/\Gamma \rightarrow 1$. As explained in the next
subsection, this implies a vanishing transmission throughout the
whole pumping cycle, in accordance with the conclusions of
Refs.~\onlinecite{Levinson02PhA,Alekseev02}.

\subsection{Relation to conductance}\label{SecPumpCond}
The criteria for the validity of the resonance approximation,
listed in the previous subsection, can be \emph{quantitatively}
checked in experiments (or in numerical calculations) by
monitoring the conductance between different leads as a function
of parameters along the pumping contour
\cite{Levinson02PhA,Entin02res}.  A definitive signature of the
relevant transport regime (for having a significant non-zero
pumped charge) would be the presence of an even number of
well-separated peaks in the conductance time trace: each resonance
($M$ in Fig.~\ref{fig1}) is associated with two peaks in the
instantaneous transmission, encountered at the two resonance
points ($B$ and $D$) where the pumping contour crosses the
resonance line on each side of the resonance, as schematically
shown in Fig.~\ref{fig1}. Note that this measurement is
independent of time: one simply measures the conductance at
different points on the pumping contour.

The contribution of each particular conductance peak to the
pumped charge can be calculated along the following lines.
Application of the general expression of the transmission
probability\cite{Datta97} from channel $\alpha'$ to channel
$\alpha$, $\mathcal{T}_{\alpha\alpha'}   = - \int dE \, f'(E) \trace
[ G_d^{\dagger} \hat\Gamma_{\alpha}
  G_d \hat\Gamma_{\alpha'} ]$, to our resonance model (as defined
  in Eq. \eqref{HdotRes}) gives the standard Breit-Wigner\cite{Breit36}
result (see e.g. Ref.~\onlinecite{Alhassid00}):
\begin{align} \label{BWTransm}
  \mathcal{T}_{\alpha\alpha'} & = - \int dE \,  f'(E)  \frac{\Gamma_{\alpha'} \Gamma_{\alpha}}{(E-E_0)^2 + (\Gamma/2)^2} \, .
\end{align}

Let us consider for simplicity an example of $L$  single-mode
leads. By using the multi-terminal Landauer conductance
formula\cite{Buttiker86} for spinless electrons,
$\mathcal{G}_{\alpha\alpha'}=(e^2/h) \mathcal{T}_{\alpha\alpha'}$,
in Eq.~\eqref{BWTransm}, we recover
well-established\cite{Alhassid00} results for the peak conductance
of a strongly pinched QD, that are related to
Eq.~\eqref{partialQres} in  an extremely simple way:
\begin{align}
  \mathcal{G}_{\alpha\alpha'}^{\text{peak}}  & = \frac{e^2}{h}
  \frac{4 \Gamma_\alpha  \Gamma_{\alpha'}}{\Gamma D} \equiv \frac{4 \Gamma}{h D} \Delta
  Q^\text{res}_{\alpha} \Delta Q^\text{res}_{\alpha'},
\label{GandQres}
\end{align}
where
\begin{equation}
  D  = \left \{ \begin{array}{cl}
                          \Gamma  , & kT \ll \Gamma , \\
                          (8/\pi) k T  , & kT \gg \Gamma .
                          \end{array}
                          \right .
  \label{GandQres1}
\end{equation}

Measurements of the peak conductance at a particular resonance
point for fixed temperature and all possible  combinations of
source and drain leads would give, in principle,  $(L^2-L)/2$
experimental values to be used in Eqs.~\eqref{GandQres}. Together
with Eq.~\eqref{totalQres},  this gives $(L^2-L)/2+1$
equations for  the $L+1$ unknowns $\Delta
Q^\text{res}_{\alpha}$ and $\Gamma/D$. Measurement of  the
temperature dependence of $\mathcal{G}^\text{peak}_{\alpha\alpha'}(T)$
would yield $D(T)$, and thus
determine $\Gamma$.  We see that even for $L=2$ it is possible to
predict the adiabatically pumped charge from the conductance
measurements, and for $L>2$ different cross-checks become
feasible.

Additional input of a few bits of information is necessary to make
the solution of Eqs.~\eqref{GandQres} and \eqref{totalQres}
unique. For a specific resonance `res', all the charges
$\Delta Q_\alpha^\text{res}$ (for all $\alpha$)  have the same
sign, determined by the type of the resonance: ``$+$'' for
loading and ``$-$'' for unloading,  see Eq.~\eqref{partialQres}.
An additional sign uncertainty arises in the
case of
two terminals ($\alpha=l,r$):
the respective
equation for the pumped charge,
 $\Delta Q_l (e- \Delta Q_r) = \mathcal{G}_{lr}^{\text{peak}} (hD/4\Gamma)$,
is symmetric
under inversion, $l \leftrightarrow r$. The resolution of these
uncertainties depends on  the particular experimental
situation, and should be easy in simple cases. We illustrate this
point  in Sec.~\ref{secEx} below, when we discuss a
two-terminal example.

\subsection{Adiabaticity condition}
One condition for the validity of the adiabatic picture requires
that an electron should have enough time to tunnel under the
barriers while the system is at resonance. Thus, the inverse
tunnelling rate  $\hbar/\Gamma$  should be much smaller than the
duration of the resonance, $\tau_r = D/| \dot{E}_0 |$, yielding the
adiabaticity condition,
\begin{align}\label{adcond}
  \hbar | \dot{E}_0 | \ll \Gamma \, D \, .
\end{align}
This condition implies that  both  the amplitude and  the
frequency of  the pumping potential must be sufficiently small for
an adiabatic pump\cite{Entin03japan}. The resonance duration
$\tau_r$ can be extracted from measurements of the conductance as
follows: measuring the variation of the conductance through the
resonance, using a very low frequency $\omega_0$, would yield the
resonance width $\tau_{r0}$ for that frequency. The value of
$\tau_r$ relevant for the pumping experiments can then be found by
rescaling, $\tau_r=\tau_{r0}\omega_0/\omega$.

At zero temperature, $D= \Gamma$ and the condition \eqref{adcond}
can be compared to the adiabaticity criterion for coherent pumping
formulated recently by Moskalets and B\"{u}ttiker
\cite{MoskaletsButtiker02}. They consider the number of side-bands
$n{_\text{max}}$ required to describe adequately the Fourier
transform of the instantaneous scattering matrix. In our case the
resonant peak of transmission in the time domain has the width
$\tau_r$, and the number of relevant Fourier harmonics
$n{_\text{max}}$ is at least $(\omega \tau_r)^{-1}$, where
$\omega$ is the cyclic frequency of the pump. The adiabaticity
criterion of Ref.~\onlinecite{MoskaletsButtiker02} states that the
scattering matrix should vary little  with energy over the range
$E \pm \hbar \omega n{_\text{max}}$. Since our characteristic
energy scale for the scattering matrix is $\Gamma$, the condition
of Ref.~\onlinecite{MoskaletsButtiker02} takes the form $\Gamma
\gg \hbar \omega n{_\text{max}}=\hbar \tau_r^{-1}$, equivalent to
Eq.~\eqref{adcond}.

For a thermally broadened resonance, the adiabaticity condition
follows also from the requirement that the first order correction
$\tilde{P}$ to the adiabatic solution of the kinetic equation
\eqref{Master} remains small: $\tilde{P} \ll 1 \Rightarrow
|f'(E_0) \dot{E_0}| \ll \Gamma/\hbar \Rightarrow $
Eq.~\eqref{adcond}.

\subsection{Application to complicated pumping potentials}\label{SecAppl}
In the resonance approximation, the pumped charge is expressed in
terms of the resonance \emph{points}, where the pumping contour
crosses the resonance lines, and  do not require the full
information on the contour in the parameter space.  We now discuss
the conditions under which Eq.~\eqref{partialQres} can be used to
obtain efficient approximate estimates of the pumped charge for a
model Hamiltonian $\hamsup{d}$, which is complicated enough to
render an exact integration\cite{Aharony02PRL} of
Eq.~\eqref{currentGeneral} impractical. Even when the validity of
the resonance approximation is marginal, such an approximate
estimate could provide a handy tool for exploring complicated
pumping models (e.g.
Ref.~\onlinecite{Levinson02PhA,Entin02res,Aharony02PRL}) and
identifying the relevant physical parameters. For simplicity, we
restrict this discussion to zero temperature.

To leading order in the coupling strengths $\lambda_\alpha$, the
parameters of the resonant level in Eq.~\eqref{HdotRes} are given
by the eigenstate of the decoupled $\hamsup{d}(t)$ which is the
closest to the Fermi energy $\mu$. Therefore, the following
algorithm can be formulated:
\begin{enumerate}
  \item Diagonalize $\hamsup{d}(t)$ (analytically or numerically),
  to get the spectrum $\{ \epsilon_m(t), \ket{\psi_m(t)} \}$.

  \item Calculate the time-dependent decay widths
  $\Gamma^{m}_{\alpha}(t)=
  \melement{\psi_m(t)}{\hat{\Gamma}_\alpha(E=\mu)}{\psi_m(t)}$ and
  shifted energy levels $ \epsilon'_m(t) =\epsilon_m(t) +
  \melement{\psi_m(t)}{\hat{\enshift}(E=\mu)}{\psi_m(t)}$.

  \item For every $m$, find all such times $t_{m,j}$ when the
  resonance condition $\epsilon'_m(t_{m,j})=\mu$ is satisfied.

  \item At each resonance time $t=t_{m,j}$,
  compute the  corresponding partial charge
  $q^{m,j}_\alpha = e \Gamma_\alpha^{m}/\sum_{\alpha'} \Gamma_{\alpha'}^{m}$.

  \item Calculate the total pumped charge as
  \begin{equation} \label{Qtot}
     Q^{\text{R}}_{\alpha} =-\sum_{m,j} q^{m,j}_\alpha \sgn \dot{\epsilon}'_m(t_{m,j}) \, ,
   \end{equation}
   or set $Q^{\text{R}}_{\alpha}=0$ if no resonances were found in step~3.
\end{enumerate}
The application of this algorithm is justified under the
conditions listed in the beginning of this section. The most
important condition is the consistency of the perturbation
expansion, $\Gamma_\alpha^{m}(t_{m,j}) \ll \Delta(t_{m,j})$, where
$\Delta(t)$ is the level spacing of $\hamsup{d}(t)$ at the Fermi
surface.

The algorithm will fail for certain values of the adjustable (not
pumping) parameters of the model, for which the number of
resonance points found in step~3 changes. This change corresponds
to  the appearance (or annihilation) of a pair of
loading/unloading resonances. Such a crossover is usually
manifested by a sharp change (a
step)\cite{Entin02res,Aharony02PRL} in the total pumped charge, as
function of the model parameters.

\section{Example: turnstile model}\label{secEx}
We illustrate the resonance approximation  by a simple example of
a single energy level with adiabatically varying couplings to the
left and right reservoirs (single level turnstile
model\cite{Entin02form}). Applications to more complicated models,
such as pumping by surface acoustic waves\cite{Aharony02PRL}, will
be reported elsewhere.

\subsection{The turnstile pumping model}
The single level turnstile  model, discussed in
Ref.~\onlinecite{Entin02form}, can be described as a special case
of the general Hamiltonian \eqref{labelHamfull}, with $N=1$ site
(and a single energy $h_{11}=\epsilon$) on the QD and with $L=2$
leads, denoted by $\alpha=l,r$.
It is now convenient to use a slightly different notation:
Consider an infinite chain of tight
binding (TB) sites, enumerated by $n=0, \pm1, \ldots$. The
site $n=0$, which represents the QD, has a
 time-independent energy $\epsilon$ and defines
$\hamsup{d}_\text{res} = \epsilon \, d^\dagger d$, with
eigenstate $|\psi \rangle$. The sites with $n>0$ ($n<0$) form the
right (left) single-mode TB lead:
\begin{align}\label{TBleads}
   \hamsup{l}_{\alpha}  & = -\sum_{n=\pm 1}^{\pm\infty} J \left(
   c^\dagger_{n} c_{n\pm1} + c^\dagger_{n\pm1} c_{n} \right) \, ,
\end{align}
where the upper sign refers to $\alpha=r$. The coupling operators
are $L_\alpha = \sqrt{X_\alpha(t)} J \,  c_{\pm1}^\dagger d$, with
the two time-dependent pumping parameters
$X_{\alpha}=\lambda_\alpha^2$.

The Hamiltonian of the leads \eqref{TBleads} is characterized by
the dispersion relation $E_{\myvec{k}}=-2 J \, \cos ka$ and the
retarded  Green's function
\begin{align}
  [G_\alpha^l ]_{nm} & =  \frac{e^{i ka |n-m|} - e^{ i ka | m+n|}}{i 2 J \sin ka} \, ,
\end{align}
where $a$ is the nearest-neighbor distance. The self-energy
operator [Eq.~\eqref{selfenDef}] is $\hat\selfen  = - (X_l+X_r) \,
J \, e^{ika} \, d^\dagger d$.

We consider the zero temperature limit and parametrize the on-site
energy as $\epsilon = (-2+\delta) J \cos \kappa a $, where the
dimensionless parameter $\delta$ is a measure of the de-tuning of
the isolated level $\epsilon$ from the Fermi energy $\mu = -2 J
\cos \kappa a$ in the leads. Near the band bottom one has $\delta
\approx (\epsilon - \mu)/J$.

\begin{figure}[tb]
  \includegraphics[clip=true,width=8cm]{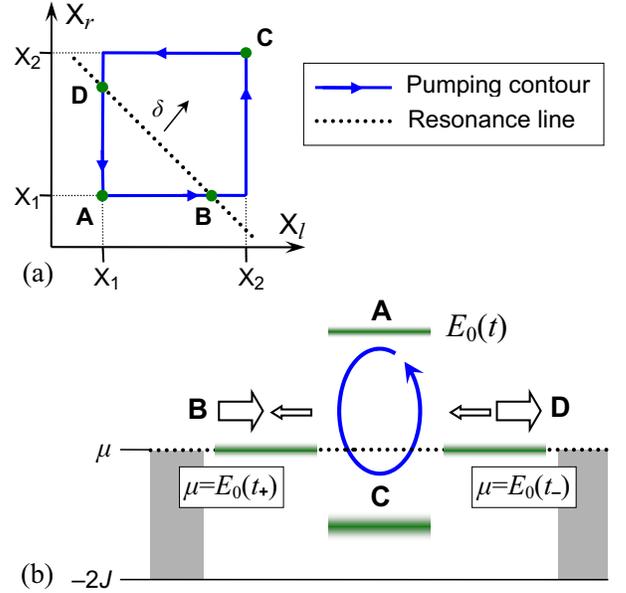}
  \caption{(color online)
  (a) The pumping contour A-B-C-D-A and the resonance
  line B-D for the single level turnstile model\cite{Entin02form}.
  (b) Interpretation of the pumping cycle on an energy diagram. A:
  The effective energy level $E_0$ is above the chemical potential
  $\mu$, the dot is empty. B: Loading process with preference to
  the left-coming electrons. C: The level $E_0$ is below $\mu$,
  the dot is occupied. D: Unloading process with preference to the
  right-going electrons. The asymmetry between $B$ and $D$ creates the
  non-vanishing total pumped charge. The arrows indicate
  schematically  the direction and  the relative magnitude of the current
  pulses caused by
  each resonance.} \label{fig2}
\end{figure}

The resonance parameters at the Fermi surface are
\begin{align}\label{resParams}
  \Gamma_\alpha & = 2X_\alpha \, J \, \sin \kappa a \, , \nonumber\\
  E_0 & =  (-2+\delta -X_l - X_r) J \cos \kappa a \, .
\end{align}
Both $\Gamma_\alpha$ and $E_0$ depend on time via the
time-dependent
couplings $X_\alpha$, which span the parameter space $\{ X_l,~X_r
\}$. The resonance condition $E_0=\mu$ defines the resonance line,
$X_l+X_r=\delta$. For an explicit calculation, we next choose the
pumping contour to be a square with corners at points $A(X_1;
X_1)$ and $C(X_2; X_2)$, as used in Ref.~\onlinecite{Entin02form}
(This is shown in Fig.~\ref{fig2}a, which forms an explicit
example of Fig.~\ref{fig1}).

The necessary conditions of Sec.~\ref{SecQuantBig}, for having
distinct resonances, are satisfied only at the bottom of the TB
band ($\sin \kappa a \ll \cos \kappa a$). As we gradually increase
$\delta$ from zero, the resonance line in Fig.~\ref{fig2}a moves
in the direction indicated by the small arrow. The resonance line
crosses the contour only if $2 X_1 \equiv \delta_1 < \delta <
\delta_3 \equiv 2 X_2$. Therefore, within the resonance
approximation we have
\begin{align}\label{QR1}
  Q^\text{R}/e = 0, \ \ \text{ if } \delta < \delta_1 \text{ or } \delta >
  \delta_3.
\end{align}

For the direction of the contour
shown  by the arrows in Fig.~\ref{fig2}, the resonance point $B$ corresponds to
loading of the dot
mostly from the left ($\Gamma_l > \Gamma_r$).
Its complementary resonance point $D$ is associated with
unloading mostly to the right ($\Gamma_l < \Gamma_r$). This
interpretation is illustrated in Fig.~\ref{fig2}b.

 At the lower left part of the contour, $\delta <
\delta_2=(\delta_1+\delta_3)/2=X_1+X_2$, the resonance points are
$D(X_1, \delta-X_1)$ and $B(\delta-X_1, X_1)$. The partial charges
pumped from the left [using Eqs.~\eqref{partialQres} and
\eqref{resParams}] are
 \begin{align}
     \Delta Q^D_l & = - e \frac{2 X_l \,  J \sin \kappa a}{2(X_l+X_r) \,  J \sin \kappa a}
     = - e \frac{X_1}{\delta}, \label{Qbd}\\
     \Delta Q^B_l & = e \frac{\delta - X_1}{\delta},
 \end{align}
where we have used $\sgn \dot E_0=\sgn
\frac{d}{dt}(-X_l-X_r)=+1$ for point $D$. The net pumped charge is
thus
\begin{align}\label{QR2}
  Q^\text{R}/e &= Q^\text{R}_l/e=(\Delta Q^D_l+\Delta Q^B_l)/e=-Q^\text{R}_r/e \nonumber\\
  &=1-(\delta_1/\delta)  ,  \ \ \text{ if } \delta_1 < \delta <
  \delta_2.
\end{align}
A similar analysis for crossing at $D(\delta-X_2, X_2)$  and
$B(X_2, \delta-X_2)$ (when $\delta_2<\delta<\delta_3$) yields
\begin{align}\label{QR3}
      Q^{\text{R}}/e= (\delta_3/\delta)-1  ,  \ \ \text{ if } \delta_2 < \delta <
      \delta_3.
\end{align}

Our resonance approximation results for $Q^\text{R}/e$ are shown
for some typical parameters (together with the exact results, see
below) in Fig.~\ref{fig3}. These results agree qualitatively with
those of Refs.~\onlinecite{Levinson02PhA} and
\onlinecite{Entin02res}: $Q^{\text{R}}/e$ reaches its maximum
value $(X_2-X_1)/(X_1+X_2)$ at $\delta=\delta_2$, where the
resonance points $B$ and $D$ are farthest away from the resonance
point $M$, which occurs at $X_1=X_2=\delta/2$. Note that
$Q^\text{R}/e$ approaches the quantized value 1 when $X_2/X_1 \to
\infty$, i.e. when the transmission at the resonance points
(related to $4X_1X_2/(X_1+X_2)^2$, via Eq. \eqref{BWTransm})
vanishes. This is consistent with Ref.~\onlinecite{Levinson02PhA},
which required that ``a large part of the resonance line'' be
surrounded by the pumping contour.

\subsection{Comparison with exact results}

The formula \eqref{ResonanceCurrent} for the resonance current is
exact in our case. Substitution of Eq.~\eqref{resParams} into
Eq.~\eqref{ResonanceCurrent} and integration over the contour
$A$-$B$-$C$-$D$ gives the total pumped charge in the form
\begin{equation} \label{turnstileQ}
  Q= \frac{e}{\pi} \int dX \left[ F(X, X_1) - F(X, X_2) \right] \, ,
\end{equation}
where
\begin{equation}
  F(X,Z)=\frac{(\delta -2 \, Z) \sin \kappa a \, \cos \kappa a}
  {(\delta - X -Z)^2 \cos^2 \kappa a + (X+Z)^2 \sin^2 \kappa a} \, .
\end{equation}
This result was obtained in Ref.~\onlinecite{Entin02form} using
the time-derivatives of the scattering matrix.

\begin{figure}[tb]
  \includegraphics[clip=true,width=8cm]{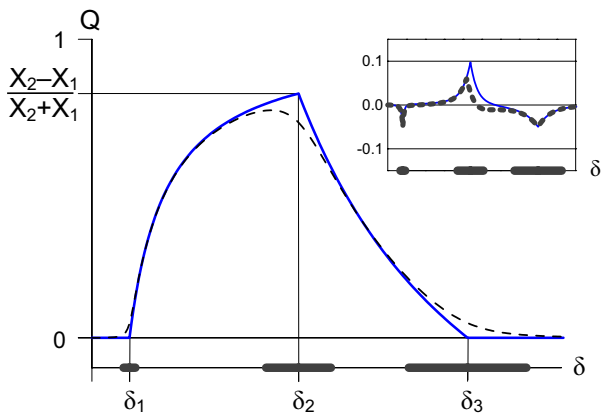}
  \caption{(color online) Pumped charge (in units of $e$)
  as a function of $\delta$  for
  $X_1=1/50$, $X_2=1/5$  and $\kappa a = \pi /20$, calculated within
  the resonance approximation ($Q^{\text{R}}$, blue continuous line) and
  exactly ($Q$, dashed line).
  Thick bars on
  the $\delta$ axis mark the resonance
  widths $\pm \Gamma/J$  around
  the special points $\delta_{1,2,3}$, where deviations from
  the the exact result are anticipated.
  Inset: Absolute error of the resonance approximation,
  $(Q^{\text{R}}-Q)/e$,
  for the same values of $\delta$. The thick dotted line corresponds to
  $Q^{\text{R}}$ calculated from the transmission maxima,
  see text for details.} \label{fig3}
\end{figure}

In Fig.~\ref{fig3} we compare the exact $Q$ and the approximate
$Q^\text{R}$.
As the resonance line in Fig.~\ref{fig2} moves from point $A$
to $C$, the pumped charge  rises from zero to a maximum, close to
$(X_2-X_1)/(X_2+X_1)$, and then falls back towards zero.
 Except for the vicinity of the special points $\delta=\delta_1,~\delta_2$ and $\delta_3$,
 there is an excellent agreement between Eqs.~(\ref{QR1},\ref{QR2},\ref{QR3}) and
Eq. \eqref{turnstileQ}.

The most significant source for deviations of the exact pumped
charge $Q$ from the separated resonance result $Q^\text{R}$ is the
term proportional to $\dot{\Gamma}_\alpha$ in the expression of
the pumping current \eqref{ResonanceCurrent}:
\begin{align}\label{deviation}
  Q - Q^\text{R} \approx  \frac{e}{2 \pi}
  \int \frac{d\Gamma_\alpha \, (E_0-\mu)  }{(\mu-E_0)^2+(\Gamma/2)^2}
\end{align}
In our example, Eq. \eqref{resParams} yields
$d\Gamma_\alpha/dE_0=\dot\Gamma_\alpha/\dot E_0=-2 \tan(\kappa a)$
when the resonance is on an edge of the contour curve where only
$X_\alpha$ varies. Thus, the integral in Eq. \eqref{deviation} is
negligibly small as long as the distance between the resonance
point and a corner of the square contour is larger than
$\Gamma/J$. Indeed, this agrees with Fig.~\ref{fig3}, where the
regions $|\delta-\delta_i|<\Gamma/J$ are indicated by horizontal
bars on the $\delta$-axis.

\begin{figure}[tb]
  \includegraphics[clip=true,width=8cm]{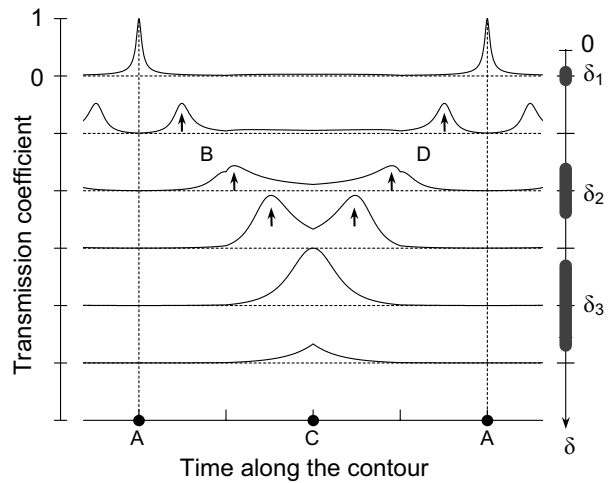}
  \caption{Time traces of the transmission coefficient
  $\mathcal{T}_{lr}$ along the pumping contour
  for six values of $\delta$, increasing with constant intervals from
  top to bottom. Two complementary resonances $B$ and $D$ (marked with arrows)
  are   observed for  $\delta_1 < \delta < \delta_3$ when the pumping contour
  crosses the resonance line (cf.~Fig.~\ref{fig2}a). Thick  bars
  on the $\delta$-axis mark the regions where the loading/unloading
   pumping mechanism fails.}
  \label{fig4}
\end{figure}
Figure \ref{fig4} shows the Breit-Wigner transmission coefficient
$\mathcal{T}_{lr}$, calculated from Eqs.~\eqref{BWTransm} and
\eqref{resParams} as a function of time (defined homogeneously
along the pumping contour) for several values of $\delta$. As
$\delta$ is increased from zero, a single peak develops at
$\delta=\delta_1$, then splits into two independent resonances $B$
and $D$, which move along the pumping contour and finally merge at
$\delta=\delta_3$ and disappear. By comparing Fig.~\ref{fig4} to
Fig.~\ref{fig3} one can follow the correlation between the
presence of  separate well-defined transmission peaks and
the validity of the resonance approximation for the pumped
charge.

\subsection{Relation to transmission}

The quantitative relation between the pumped charge and the
transmission (conductance) has been discussed in
Sec.~\ref{SecPumpCond}. In order to illustrate this discussion,
let us assume that the transmission traces (Fig.~\ref{fig4}) are
the \emph{only} available data for our two terminal system. One
observes two resonances in the range $\delta_1 < \delta <
\delta_3$ -- both giving the same value of the peak transmission
$\mathcal{T}_{\text{max}}$. One of the resonances represents
loading, contributing $\Delta Q^\text{res}_\alpha >0$, while the
other one necessarily  represents unloading (with $\Delta
Q^\text{res}_\alpha <0$). If we make a mistake at this stage and
take the wrong sign in Eq.~\eqref{totalQres}, it will  only change
the assumed pumping direction, $Q^\text{R} \to -Q^\text{R}$. Let
us treat the first resonance as loading and calculate the partial
charge pumped from the left reservoir, $\Delta Q^\text{res}_l>0$.
Solution of Eqs.~\eqref{totalQres} and \eqref{GandQres} gives two
roots, $\Delta Q^\text{res} = e (1 \pm \sqrt{1-{\cal
T}_\text{max}})/2$, and one must decide which of the two
corresponds to $\Delta Q^\text{res}_l$. The same dilemma holds for
the second resonance. Considering all four options yields three
possible answers $Q^{\text{R}}= \pm Q'$ and $0$, where $Q'=
e\sqrt{1-{\cal T}_\text{max}}$. The correct result
($Q^{\text{R}}=Q'$) may be chosen as the one which gives the best
fit to the  data of the pumping calculation/experiment. Once the
uncertain signs have been chosen correctly, there is no need to
repeat this ``trial-and-error'' procedure, since the contour
changes continuously. Of course, if some features of the pumping
contour design are known (such as which coupling is dominant in
different regions), the sign uncertainties are much easier to
resolve.

The result of the above calculation, $|Q-Q^\text{R}|/e$, is shown
in the inset of Fig.~\ref{fig3} by a thick dotted line. One can
see that both ways of calculating $Q^\text{R}$ (from the
analytic expressions (\ref{QR1},\ref{QR2},\ref{QR3}) and from
using
the peak transmission) give similar small deviations from
the exact value $Q$ of the pumped charge.

We now leave our specific example, and consider Eq.
\eqref{deviation} for a general resonance. As seen in the example,
the integral in Eq. \eqref{deviation} becomes non-zero whenever
$\gamma_\alpha=\dot\Gamma_\alpha/\dot E_0$ is not a
time-independent constant during the whole resonance. For
non-constant $\gamma_\alpha$, the largest deviation
$|Q-Q^\text{R}|$ arises when $\gamma_\alpha$ changes sign exactly
at the resonance point $E_0=\mu$; one then finds that
$|(Q-Q^\text{R})/Q^\text{R}|< x \max|\gamma_\alpha|$, where $x$ is
a number of order unity, which depends on the details of the
contour. These considerations justify Condition No. 3 in the
beginning of Sec.~\ref{SecQuantBig}.

\section{Conclusions}\label{secConc}
We have considered a general model of adiabatic quantum pumping of
spinless non-interacting electrons, in the coherent resonant
tunnelling regime. In the limit of distinct transmission
resonances along the pumping contour, the pumped charge is given
by a sum of individual contributions due to each resonance. During
each resonance one electron either enters or leaves the system,
with the probability distribution between different reservoirs
given by  the corresponding tunnelling rates
$\Gamma_\alpha/\hbar$.

We have clarified the role of quantum coherence in the
resonance-assisted pumped charge quantization by showing that
quantization arises due to population of discrete resonant states
with preference to a single reservoir in each resonance. A
quantitative and experimentally verifiable relation between the
pumped charge and the peak conductance has been proposed. The
resonance approximation also provides a simple calculational
algorithm for analyzing complex pumping potentials.

Our results remain valid if (1) the spacing $\Delta$ between
different resonant levels is much larger than $\Gamma_\alpha$,
$kT$; (2) the relative magnitude of the couplings $\Gamma_\alpha$
to different reservoirs does not change much during a resonance;
(3) the adibaticity condition $\hbar \dot{E}_0 \ll \max(\Gamma,
kT)$ is not violated.

Systematic extension of the resonance approximation
to the situations when electron-electron interactions can not be
neglected is a question of further study.

\begin{acknowledgments}
This project was carried out in a center of excellence supported
by the Israel Science Foundation.

\end{acknowledgments}

\appendix*

\section{Formula for the pumped current}
\label{A}

In this Appendix, we use standard scattering theory relations
\cite{Paulsson02,Datta97} to derive Eq.~\eqref{currentGeneral}
from Eq.~\eqref{pump_current}. The scattering states
$\ket{\chi_{\myvec{k}\alpha}}$ can be obtained from the
Lippman-Schwinger equation
\begin{align} \label{LippmanSchwinger}
  \ket{\chi_{\myvec{k}\alpha}} &= (1  + G L^{\dagger}_\alpha ) \ket{w_{\myvec{k}\alpha}} \, ,
\end{align}
where $G=(E+i 0-\ham )^{-1}$ is the retarded Green's function
taken at energy $E=E_{\myvec{k}\alpha}$. The time $t$ enters
Eq.~\eqref{LippmanSchwinger} as a parameter.

Defining projection operators $\hat P^d$ and $\hat P_\alpha$ onto
the QD and onto lead $\alpha$, one has
\begin{eqnarray}
& \hamsup{d}=\hat P^d \hamsup{d} \hat P^d, \ \ \ G_d=\hat P^d G_d
\hat P^d,\ \ \ \hat P_\alpha \ham \hat
P_{\alpha'}=\delta_{\alpha\alpha'} \ham^l_\alpha,
\nonumber\\
& L_\alpha=L_\alpha \hat P^d=\hat P_\alpha \ham \hat P^d, \ \ \
\hat P^d \ket{w_{\myvec{k}\alpha}}=0,
\end{eqnarray}
and therefore
\begin{align} \label{Mdot}
  M^d_\alpha & \equiv \melement{\chi_{\myvec{k}\alpha}}{\dot{\hamsup{d}}}{\chi_{\myvec{k}\alpha}} =
  \melement{w_{\myvec{k}\alpha}}{ L_\alpha G^\dagger_d \dot{\hamsup{d}}
  G_d L_\alpha^\dagger }{w_{\myvec{k}\alpha}}.
\end{align}

To derive Eq. \eqref{Gdot}, we start from $(E-\ham)G=I$, multiply
from the right by $\hat P^d$ and from the left by $\hat P_\alpha$
and -- using the identity $\hat P^d+\sum \hat P_\alpha=I$ --
obtain the relation $\hat P_\alpha G \hat P^d=G^l_\alpha L_\alpha
G_d$. A similar multiplication from the left by $\hat P^d$ then
yields Eq. \eqref{Gdot}, with
\begin{equation}
\hat\selfen=\sum_\alpha L^\dagger_\alpha G^l_\alpha L_\alpha,
\end{equation}
which is equivalent to Eq.~\eqref{selfenDef}.

Similarly, the time dependance of the coupling strengths
$\lambda_\alpha(t)$ contributes to the current $I_\alpha$ via the
matrix element $M^l_\alpha  =
\sum_{\alpha'}\melement{\chi_{\myvec{k}\alpha}}{\dot{L}_{\alpha'}}{\chi_{\myvec{k}\alpha}}
+ \text{h.c}$. Using the trivial relations $\dot L_\alpha=(\dot
\lambda_\alpha/\lambda_\alpha)L_\alpha$ and $\hat P_{\alpha'}
\ket{w_{\myvec{k}\alpha}}=\delta_{\alpha\alpha'}\ket{w_{\myvec{k}\alpha}}$,
a straightforward calculation gives:
\begin{align} \label{Mleads}
   M^l_\alpha= \melement{w_{\myvec{k}\alpha}}{ L_\alpha \left [
     G^\dagger_d
     \dot{\hat{\enshift}}
     G_d L_\alpha^\dagger
     + \frac{\dot{\lambda}_{\alpha}}{\lambda_{\alpha}}
     \left( G_d + G_d^{\dagger} \right )
     \right  ] L_\alpha^\dagger
     }{w_{\myvec{k}\alpha}} \, .
\end{align}

The normalization to the unit flux, $\bra{w_{\myvec{k}\alpha}}
w_{\myvec{k'}\alpha}\rangle=(2\pi/v_{k\alpha})\,\delta(k-k')$
 implies that $\hat
 P_\alpha=\int(dk/2\pi)v_{k\alpha}\ket{w_{\myvec{k}\alpha}}\bra{w_{\myvec{k}\alpha}}$. Using
 also the standard relation $i(G^l_\alpha-G^{l
 \dagger}_\alpha)=\ket{w_{\myvec{k}\alpha}}\bra{w_{\myvec{k}\alpha}}/\hbar$,
 we find the relation
\begin{align} \label{gammadensity}
  \hbar \hat{\Gamma}_{\alpha} = L_{\alpha}^{\dagger} \ket{w_{\myvec{k}\alpha}}\bra{w_{\myvec{k}\alpha}}  L_{\alpha} \, .
\end{align}
Introducing the trace over the QD's subspace, using
\eqref{gammadensity} in Eqs.~(\ref{Mdot}) and (\ref{Mleads}), and
substituting the results into ${\cal I}_\alpha  =
(M_\alpha^d+M_\alpha^l )/\hbar$, we finally end up with Eq.
\eqref{currentGeneral}.


\end{document}